\documentclass[12pt]{iopart}

\usepackage{iopams}
\usepackage{graphicx}

\begin{document}

\title{Benefits of Artificially Generated Gravity Gradients
for Interferometric Gravitational-Wave Detectors}

\author{L.~Matone$^1$, P.~Raffai$^2$, S.~M\'{a}rka$^1$, R.~Grossman$^1$,
P.~Kalmus$^1$, Z.~M\'{a}rka$^1$, J.~Rollins$^1$, V.~Sannibale$^3$}
\address{
$^1$ Columbia University, Department of Physics, New York, NY 10027, USA\\
$^2$ E\"otv\"os University, Institute of Physics, 1117 Budapest, Hungary\\
$^3$ California Institute of Technology, LIGO Laboratory, Pasadena, CA 91125, USA}

\begin{abstract}
We present an approach to experimentally evaluate gravity gradient
noise, a potentially limiting noise source in advanced
interferometric gravitational wave (GW) detectors. In addition, the
method can be used to provide sub-percent calibration in phase and
amplitude of modern interferometric GW detectors. Knowledge of
calibration to such certainties shall enhance the scientific output
of the instruments in case of an eventual detection of GWs. The
method relies on a rotating symmetrical two-body mass, a Dynamic
gravity Field Generator (DFG). The placement of the DFG in the
proximity of one of the interferometer's suspended test masses
generates a change in the local gravitational field detectable with
current interferometric GW detectors.
\end{abstract}

\maketitle

\section{Introduction}

Dynamic gravity fields generated by rotating masses have been used
previously in several experimental tests; however, their
exploitation in conjunction with interferometric gravitational
detectors has not been addressed until now. Forward and Miller
\cite{ForwardMiller1967} in 1967 developed a gravity field generator
that allowed them to calibrate an orbiter sensor capable of
measuring the lunar mass distribution. A similar technique was used
by Weber {\it et al.} \cite{SinskyWeber1967,Sinsky1968} to calibrate
a GW bar detector, where a volume of matter was acoustically
stressed at $1660\ \mathrm{Hz}$ and the resulting noise excess in
the detector was found to be consistent with theory. At the
University of Tokyo, in the 1980s, a series of experiments were
conducted to test the law of gravitation up to a distance of $10\
\mathrm{m}$ \cite{Hirakawa1980,oide,suzuki,Ogawa1982,Kuroda1985}. In
these studies, the coupling between the dynamic field, generated by
a rotating mass, and the quadrupole moment of a mechanical
oscillator antenna was measured confirming the gravitational law
within experimental uncertainties \cite{Ogawa1982,Kuroda1985}. In
the 1990s, the gravitational wave group at the University of Rome
developed and carried out experiments \cite{Astone1991,Astone1998}
on the cryogenic GW bar detector, EXPLORER, at CERN. A device, with
quadrupole moment of $\mathcal{M}_\mathrm{2} = 6.65 \times 10^{-2}\
\mathrm{kg\ m}^2$ and rotating in the frequency range of $450-470\
\mathrm{Hz}$, was developed to calibrate the antenna and was also
used to confirm existing upper limits to Yukawa-like gravitational
potential violations at laboratory scale.

The increased sensitivity and bandwidth of modern interferometric
gravitational wave detectors warrants a new investigation into and
opens exciting new possibilities for application of advanced gravity
field generators in GW research. Presently interferometric
gravitational wave detectors are reaching their design sensitivity
enabling us to probe for gravitational radiation from sources well
beyond the Local Group of galaxies. The response of these detectors
to GW radiation is usually evaluated by direct injection of possible
waveforms with known amplitude via magnetic actuators, also used for
active control of the test masses' (essentially the interferometer
mirrors) displacement. In addition, displacement in the test mass
position can be induced by local gravity fields produced by a
Dynamic gravity Field Generator (DFG). A DFG is essentially a
symmetric rotating object with a significant quadrupole moment. When
it is placed in the proximity of one of the interferometer mirrors,
the induced change due to the device's quadrupole moment can be
measured by the GW detectors such as the Laser Interferometer
Gravitational Wave Observatory (LIGO)
\cite{LIGO_NIM_2004,LIGOstatus}, the VIRGO experiment \cite{virgo},
the $300\ \mathrm{m}$ Laser Interferometer Gravitational Wave
Antenna (TAMA300) \cite{tama} and the GEO600 interferometer
\cite{geo}. Future detectors, such as Advanced LIGO (AdLIGO)
\cite{adligohomepage}, offer higher sensitivity.

Several authors (see for example
\cite{SaulsonNoise,Thorne1998,CellaNewtNoise}) pointed out that
gravity gradient (or Newtonian) noise, generated by density
fluctuations in the Earth and the atmosphere, can be a potentially
limiting noise source in advanced interferometric GW detectors.
Motion of massive bodies (e.g. due to human activity) in the
vicinity of the interferometer test masses also alters the local
gravitational field, mainly at low
frequencies~\cite{SperoNewtNoise,JapanNewtNoise,SaulsonNoise}.
Gravity gradient noise manifests itself as an induced motion of the
interferometer mirrors due to the fluctuation of the local gravity
field. The DFGs described here can be used to modulate the local
gravitational field around the test mass (TM) at a precise frequency
and phase on a well-controlled manner and thereby directly
validate/evaluate the expected noise generation and coupling
mechanisms to complex structures.

In addition, DFGs have the potential to provide sub-percent
amplitude and phase calibration of interferometric GW detectors. In
the case of LIGO, currently there are two calibration methods in
use. The first one uses the interferometer TM's coil-magnet actuator
to calibrate the gravitational wave channel (see for example
\cite{cal1} and \cite{cal2}) while the second method uses the
radiation pressure exerted on the TM by an independent laser source
(see for example \cite{cal3}, \cite{Photoncal1}, \cite{Photoncal2}
and recently \cite{Photoncal3}). A DFG provides an alternative and
independent sub-percent calibration, significantly improving the
current accuracy of several percents (see e.g. \cite{CalReview}).

In this work we describe a hypothetical two-body DFG coupled to an
ideal interferometric gravitational wave detector. The induced
displacement on the suspended TM is dominated by the quadrupole
moment of the DFG mass distribution in the case of a symmetric
device. Any undesired system asymmetry will contribute to the dipole
moment and can be measured and accounted for directly. We asses the
application of such devices for the calibration of interferometric
GW detectors as well their possible usage in gravity gradient noise
studies that will eventually limit the performance of long baseline
detectors at low frequencies.

Additionally, two DFGs in a null experiment setup can be used to
explore violations to Newton's $1/r^2$ law well beyond the current
limits. We investigated this possibility in detail for LIGO,
Advanced LIGO and Virgo detectors via numerical simulations. This is
the subject of a separate publication.~\cite{rotor}

\section{Newtonian field dynamics from a two-mass DFG}\label{NFD}

\begin{figure}
\begin{center}
  \includegraphics[width=5in]{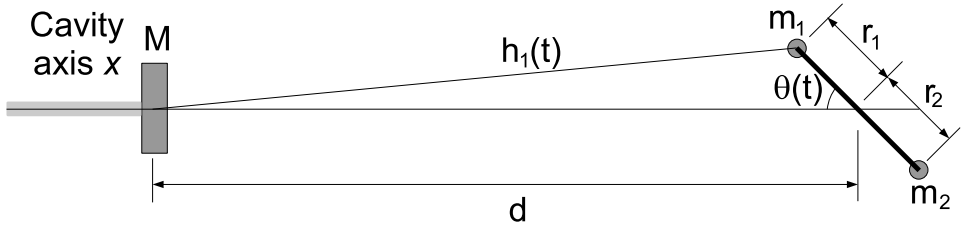}\\
  \caption{Schematic of an ideal symmetric two-mass DFG.
  The system consists of two masses, $m_\mathrm{1}$ and $m_\mathrm{2}$, separated
  by a distance of $r_\mathrm{1}$ and $r_\mathrm{2}$ from the center of rotation.
  The center of rotation is placed a distance $d$ away from the
  mirror's center of mass. The system rotates at a frequency of $f_\mathrm{0} = \omega_\mathrm{0}/(2 \pi)$
  where $\theta(t)=\omega_\mathrm{0} t$. The $x$ axis denotes the
  interferometer's optical axis and only accelerations along this axis are considered.}
  \label{ggm_diagram}
\end{center}
\end{figure}

In analytical derivations, throughout this paper we will treat the
suspended interferometer TM and the masses of the two-body DFG as
point masses for simplicity. First we calculate the acceleration,
along the laser beam axis, the mass is subjected to from a DFG
configuration shown in fig.(\ref{ggm_diagram}). Masses
$m_\mathrm{1}$ and $m_\mathrm{2}$ are separated by a distance
$r_\mathrm{1}$ and $r_\mathrm{2}$, respectively, from the center of
rotation and are rotating at a frequency of $f_\mathrm{0} =
\omega_\mathrm{0}/(2 \pi)$. The center of mass of mirror $M$ and the
DFG's center of rotation are separated by a distance $d$, where $d >
r_\mathrm{1,2}$.

Assuming that the distance between the DFG's $i$-th mass and the
mirror is $h_\mathrm{i}$, the Newtonian potential at the mirror's
center of mass is
\begin{equation}
V^\mathrm{c}  = \sum\limits_{i = 1}^2 {V_\mathrm{i}^\mathrm{c} }  =
- GM\sum\limits_{i = 1}^2 {\frac{{m_\mathrm{i} }} {{h_\mathrm{i}}}}.
\end{equation}
Introducing the variables $R_\mathrm{1} = r_\mathrm{1}/d$, and
$R_\mathrm{2} = -r_\mathrm{2}/d$, $h_\mathrm{i}$, being a function
of time can be written as
\begin{equation}\label{hi}
h_\mathrm{i}(t)  = d ~ \sqrt {1 + R_\mathrm{i}^2  - 2R_\mathrm{i}
\cos \theta(t) }
\end{equation}
where $\theta(t)=\omega_\mathrm{0} t$ (see fig.(\ref{ggm_diagram})).
The magnitude of the TM's induced acceleration along the laser beam
axis is
\begin{equation}\label{cla_acc}
a^\mathrm{c}  = \frac{1} {M}\left| {\frac{{\partial V^\mathrm{c} }}
{{\partial d}}} \right| = \frac{G} {{d^2 }}\sum\limits_{i = 1}^2
{m_\mathrm{i} B_\mathrm{i} (R_\mathrm{i} ,\theta )}.
\end{equation}
Here $B_\mathrm{i} (R_\mathrm{i} ,\theta )$ is a geometrical factor
\begin{equation}
B_\mathrm{i} (R_\mathrm{i} ,\theta ) = \frac{{1 - R_\mathrm{i} \cos \theta }} {{\left( {1 -
2R_\mathrm{i} \cos \theta  + R_\mathrm{i}^2 } \right)^{{3 \mathord{\left/
 {\vphantom {3 2}} \right.
 \kern-\nulldelimiterspace} 2}} }}.
\end{equation}
For the case of a much smaller lever arm $r_\mathrm{i}$ than the
distance $d$ ($R_\mathrm{i} \ll 1$) we can expand $V^\mathrm{c}$
thereby expressing the induced acceleration $a^\mathrm{c}$ in terms
of the $n$-th multipole moment $\mathcal{M}_\mathrm{n}$ of the DFG's
mass distribution
\begin{equation}\label{acceleration}
a^\mathrm{c} = \frac{G}{d^2}\sum_{n=0}^\infty ~ \frac{n+1}{d^n} \cdot
\mathcal{M}_\mathrm{n} \cdot P_\mathrm{n}(\cos{\theta})
\end{equation}
where
\begin{equation}\label{Mn classical}
\mathcal{M}_\mathrm{n} = m_\mathrm{1} r_\mathrm{1}^n + (-1)^n
m_\mathrm{2} r_\mathrm{2}^n
\end{equation}
and $P_\mathrm{n}(\cos{\theta})$ is the Legendre polynomial of $n$-th order.

We remark that the DFG's dipole moment, as well as the higher-order
odd moments, contribute only to the odd harmonic terms, whereas the
quadrupole moment and the higher-order even terms, contribute only
to the even harmonic terms. In the case of an ideally symmetric DFG,
all odd moments vanish and the induced displacement is dominated by
the quadrupole moment $\mathcal{M}_\mathrm{2}$ at twice the rotation
frequency.

\subsection{Induced Displacement from the Newtonian Potential}
\label{idis}

The suspended TM can be considered as a free body for frequencies
well above the eigenfrequencies of the suspension which typically
lie around $1\ \mathrm{Hz}$ \cite{TMsuspension}. Neglecting the
time-independent term, double integrating eq.(\ref{acceleration})
with respect to time and considering only the dominant terms in the
first few harmonics, the TM's displacement along the laser beam
axis, $x$, can be written as
\begin{eqnarray}\label{displ2}
x(t) \simeq \frac{G}{(d ~ \omega_\mathrm{0} )^2} \times \Bigg{[} 2
\cdot \frac{\mathcal{M}_\mathrm{1}}{d} \cdot \cos{\omega_\mathrm{0}
t} &+&\\ \nonumber & & \hskip -1.5in \frac{9}{16} \cdot
\frac{\mathcal{M}_\mathrm{2}}{d^2} \cdot \cos{2 \omega_\mathrm{0} t}
+  \frac{5}{18} \cdot \frac{\mathcal{M}_\mathrm{3}}{d^3} \cdot
\cos{3 \omega_\mathrm{0} t} \Bigg{]}
\end{eqnarray}

In the case of a symmetric two-mass DFG, the dipole and the octopole
contribution vanishes and the quadrupole moment
$\mathcal{M}_\mathrm{2}$ dominates. For initial LIGO throughout the
paper we will consider the case of $m_\mathrm{1}=m_\mathrm{2}=1.5\
\mathrm{kg}$, $r_\mathrm{1}=r_\mathrm{2}=0.25\ \mathrm{m}$
(equivalent to a quadrupole moment of
$\mathcal{M}_\mathrm{2}=0.1875\ \mathrm{kg}\ \mathrm{m}^2$), with a
rotation frequency of $f_\mathrm{0}=\omega_\mathrm{0}/(2 \pi)=51\
\mathrm{Hz}$ and a distance of $d=2.5\ \mathrm{m}$. The resulting
RMS displacement change $x_\mathrm{rms}$ at twice the rotation
frequency is $1.24 \times 10^{-18}\ \mathrm{m}$ and scales according
to
\begin{eqnarray}
\label{displ1} x_\mathrm{rms} \simeq 1.24 \times 10^{-18}\
\mathrm{m} &\times&
\\ \nonumber && \hskip -1.5in \Bigg{(}
\frac{\mathcal{M}_\mathrm{2}}{0.1875\ \mathrm{kg}\ \mathrm{m}^2}
\Bigg{)} \times \Bigg{(} \frac{51\ \mathrm{Hz}}{f_\mathrm{0}}
\Bigg{)}^2 \times \Bigg{(} \frac{2.5\ \mathrm{m}}{d} \Bigg{)}^4
\end{eqnarray}

\begin{figure}
\begin{center}
  \includegraphics[width=5in]{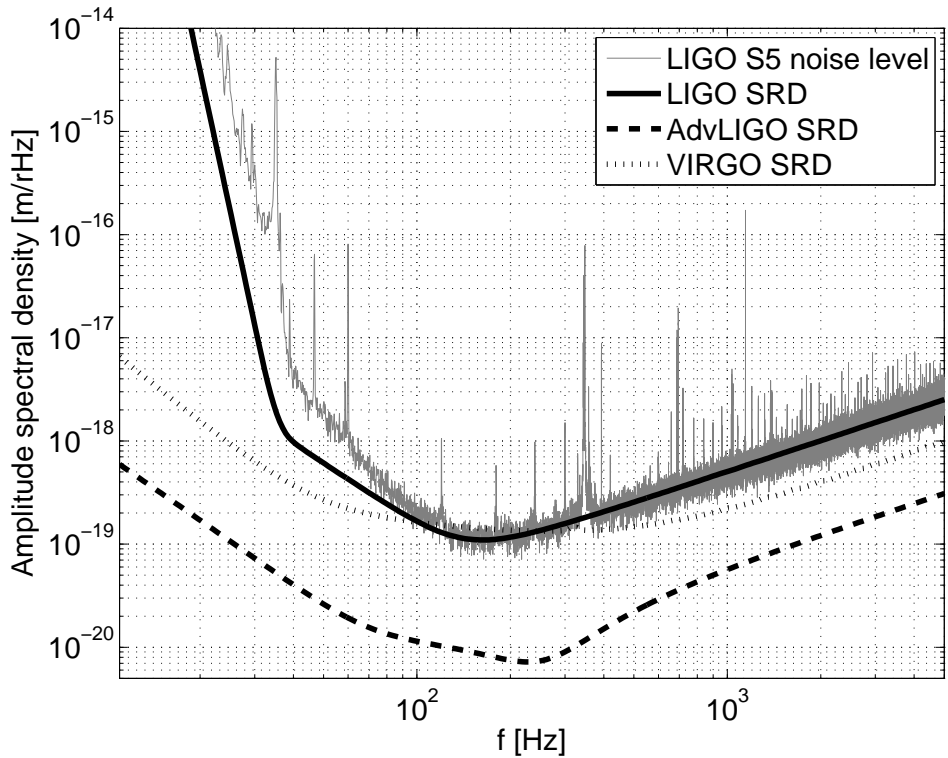}\\
  \caption{The nominal displacement sensitivity of LIGO (gray trace) at
  the beginning of the fifth science run (S5) together with its design curve
  (black); the design sensitivity (SRD) for Advanced LIGO (dashed) and the
  European VIRGO detector (dotted) are also shown.}
  \label{S5sensitivity}
\end{center}
\end{figure}

Fig.(\ref{S5sensitivity}) shows the design sensitivities for initial
LIGO, AdLIGO and VIRGO also including LIGO's nominal displacement
sensitivity for the beginning of the fifth science run
\cite{S5sensitivity} (S5). The LIGO detectors' displacement
sensitivity at $102\ \mathrm{Hz}$ is $\sim \hskip -1mm 2\times
10^{-19}\ \mathrm{m}/\sqrt{\mathrm{Hz}}$ (see gray curve in
fig.(\ref{S5sensitivity})).

The signal-to-noise ratio (SNR), defined as the ratio of the RMS
signal to the displacement noise spectrum density integrated for a
time T, gives a measure of how much a given stimulus is above
background. For the above mentioned device, in the case of LIGO
during S5 at $102\ \mathrm{Hz}$ (that is twice the above mentioned
rotation frequency), and for an integration time of $1\ \mathrm{s}$,
we obtain an SNR of 6. In general terms, for an arbitrary noise
floor $\tilde{n}$, and integration time $T$, the SNR scales as
\begin{eqnarray}\label{snr1}
SNR \simeq 6 \times \Bigg{(} \frac{2 \times 10^{-19}\
\mathrm{m}/\sqrt{\mathrm{Hz}}}{\tilde{n}} \Bigg{)} &\times& \\
\nonumber && \hskip -1.in \times \Bigg{(} \frac{T}{1\ \mathrm{s}}
\Bigg{)}^{1/2} \times \Bigg{(} \frac{x_\mathrm{rms}}{1.24 \times
10^{-18}\ \mathrm{m}} \Bigg{)}
\end{eqnarray}
where $x_\mathrm{rms}$ is shown in eq.(\ref{displ1}). At the present
sensitivity level of LIGO it is possible to sense such a dynamically
changing gravity field from the DFG in question using a relatively
small integration time.

\begin{figure}
\begin{center}
  \includegraphics[width=5in]{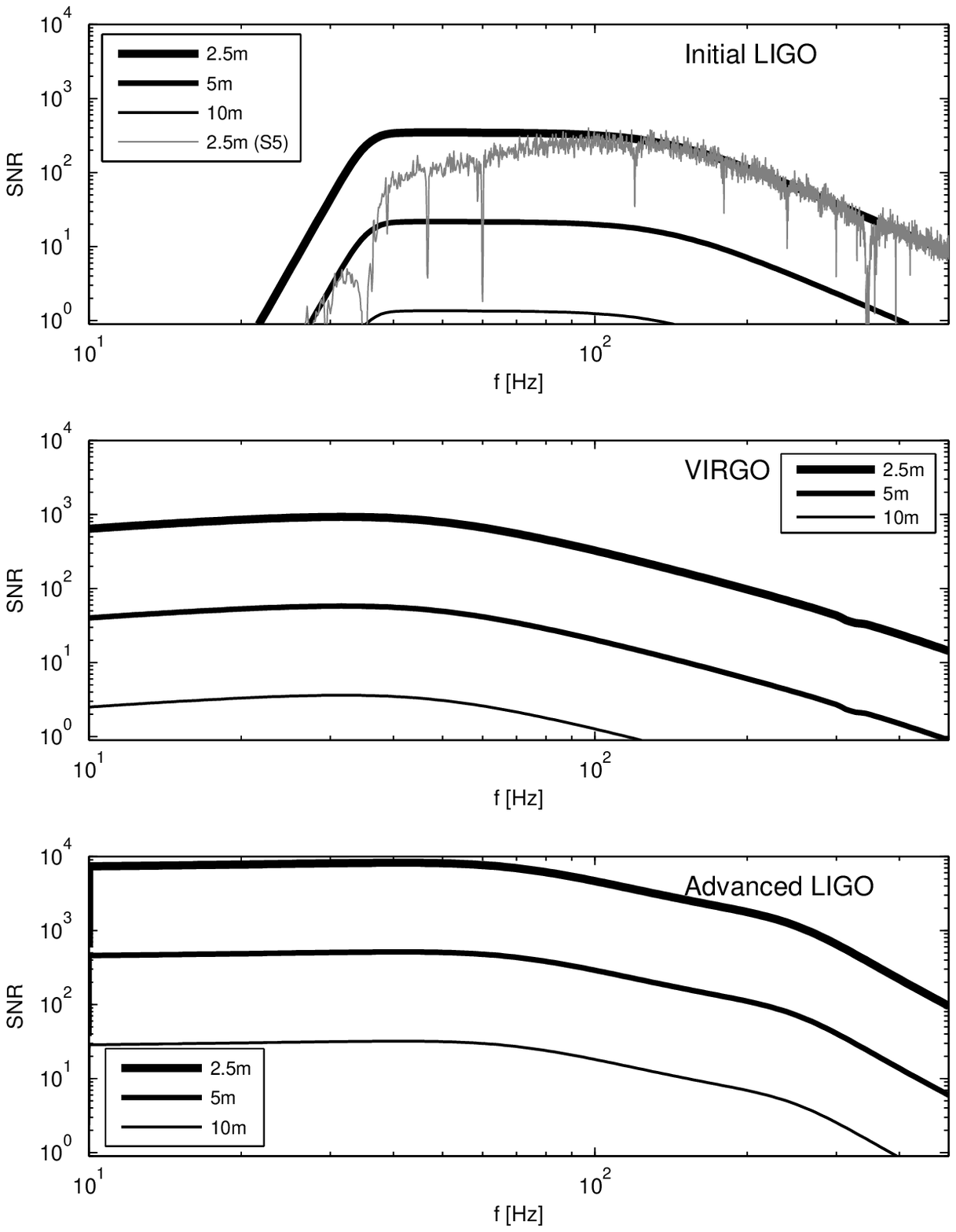}\\
  \caption{The Signal-to-Noise Ratio due to a DFG of quadrupole moment
  $\mathcal{M}_\mathrm{2}=0.1875\ \mathrm{kg}\ \mathrm{m}^2$ and 1/2 hour of integration time. Top:
  initial LIGO with the DFG positioned 2.5, 5 and 10m away from the TM; Middle: VIRGO for positions of 2.5, 5, and 10m; Bottom: Advanced LIGO
  with distances to the TM of 2.5, 5 and 10m.}
  \label{snr}
\end{center}
\end{figure}

Fig.(\ref{snr}) shows the SNR for different detectors as a function
of twice the rotational frequency with an integration time of half
an hour. Using once again the example cited above (DFG of quadrupole
moment $\mathcal{M}_\mathrm{2}=0.1875\ \mathrm{kg}\ \mathrm{m}^2$),
the top portion of the figure shows, that for rotational frequencies
ranging between $10\ \mathrm{Hz}$ and $500\ \mathrm{Hz}$, a distance
of $2.5\ \mathrm{m}$ can be used for the initial LIGO detectors.

The center part of fig.(\ref{snr}) shows the SNR for the VIRGO
detector. Due to the detector's sensitivity at low frequencies, low
rotational frequencies, as low as $\sim 10\ \mathrm{Hz}$, could be
used. The bottom portion of fig.(\ref{snr}) shows the response from
the Advanced LIGO interferometer.

\begin{figure}
\begin{center}
  \includegraphics[width=4in]{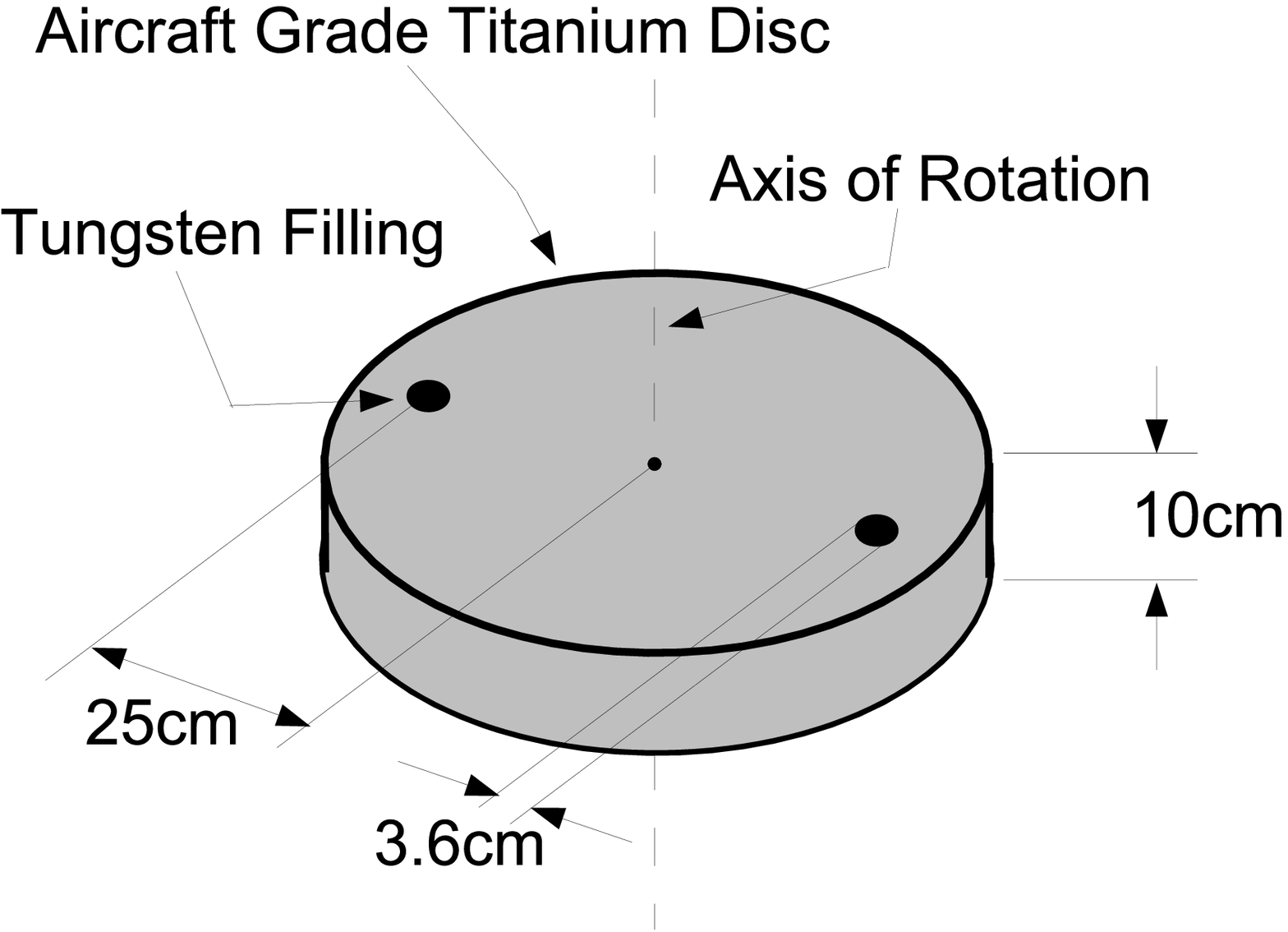}\\
  \caption{Sketch of a hypothetical DFG. The DFG consists of an
  Aircraft Grade (6Al/6V/2Sn) Titanium disc $60\ \mathrm{cm}$ in diameter and
  $10\ \mathrm{cm}$ in height. It holds two Tungsten cylinders at $25\ \mathrm{cm}$ from the
  rotation axis. The diameter of the Tungsten cylinders is $3.6\ \mathrm{cm}$.}
  \label{rotormodel}
\end{center}
\end{figure}

\subsection{A hypothetical DFG design}

In fig.(\ref{rotormodel}) we show a hypothetical DFG design based on
the parameters discussed in this section. It consists of an Aircraft
Grade (6Al/6V/2Sn) Titanium disc $60\ \mathrm{cm}$ in diameter and
$10\ \mathrm{cm}$ in height. The disc has two cylindrical slots,
$50\ \mathrm{cm}$ apart, which can hold different materials. The
choice of materials was motivated by the desire to maximize density
difference and strength while still keeping the material cost within
the bounds of reason. We use Tungsten cylinders $3.6\ \mathrm{cm}$
in diameter, corresponding to an effective mass difference of $1.5\
\mathrm{kg}$, as an example in the following sections. Practical
details, such as the expansion and stress factors of the DFG under
prolonged operating conditions must be modeled and simulated by
finite element analysis methods, then subsequently measured and
taken into account. These studies are beyond the scope of this
paper.

\subsection{Gravity Gradient Noise Studies with DFGs}

A DFG in the proximity of the test mass of the interferometer can be
used to experimentally investigate and model the coupling between
the varying gravity field and the complex suspension system of the
test mass in many fundamental configurations. The artificial gravity
gradient field generated by a DFG not only couples to the test mass
but also into all stages of the multistage suspension system and
gives rise to possible second order effects. By varying the
placement and rotation frequency of the DFG, this artificial gravity
field can simulate conceivable gravity gradient noise sources
specific to the local environment of the interferometric detector in
question. With a DFG, the dependence of the TM's displacement on the
orientation of the gravity gradient noise source can be mapped: the
DFG can be installed at different distances from the TM in the axis
of the laser beam as well as placed off-axis and out of the plane of
the interferometer. This is especially important since gravity
gradient noise couples to the system from each direction on
different ways thus potentially introduces problems into the
detection chain via hard to track second order effects and possible
nonlinear couplings. Additional advantage of artificially generated
dynamic gravity gradients is that the frequency dependence of the
interferometer's response to Newtonian noise sources could be mapped
out in detail, which is especially important for the low frequency
region. The results might eventually be used in generating
approaches for mitigating the effect of local gravity gradients in
future detectors at low frequencies besides providing accurate
information about the nature of this noise source.

\subsection{Calibration of an Interferometric GW Detector using a
DFG}\label{calibration}

In this section we address the level of precision we must achieve
when using the DFG as a calibration tool. While the present
calibration accuracy of 2-10\% in amplitude and
phase~\cite{CalReview} (depending on frequency range) might seem
adequate for upper limit and event rate studies, it will be
important to know the calibration of the detector to a higher
accuracy when the collaborations enter the "detection era".
Subpercent amplitude calibration becomes important when signals with
sufficiently large signal to noise ratios are observed. In the
context of a detected signal via a global network of interferometric
GW detectors, where the waveform is recoverable, phase calibration
known to a higher precision shall be beneficial. Coherent network
methods will perform better, pointing accuracy will increase, source
distance information can be recovered and used with a higher
accuracy. With an increase in calibration certainty the precision of
waveform and polarization recovery is expected to improve, which in
turn allows for better scientific output.

With the DFG method, the achievable calibration accuracy would be
limited by the uncertainty in the gravitational constant, G, at the
subpercent level. To estimate the calibration uncertainty we first
consider the TM displacement $x_\mathrm{rms}$ induced by the DFG due
to its quadrupole moment $\mathcal{M}_\mathrm{2}$ is given in
eq.(\ref{displ1}). In statistical terms (assuming a large number of
DFGs identical within practical tolerances), the relative
uncertainties in the measurement of the gravitational constant
($\delta G/G$), in the mass ($\delta m/m$), arm length ($\delta
r/r$), rotation frequency ($\delta f_\mathrm{0}/f_\mathrm{0}$) and
distance from the TM center of mass ($\delta d/d$) add in quadrature
leading to a relative uncertainty on the induced displacement
($\delta x/x$) and is approximately described by
\begin{eqnarray} \label{errors}
\Bigg{(} \frac{\delta x}{x} \Bigg{)}^2 \simeq \Bigg{(} \frac{\delta
G}{G} \Bigg{)}^2 + \Bigg{(} \frac{\delta m}{m} \Bigg{)}^2 + 4
\Bigg{(} \frac{\delta r}{r} \Bigg{)}^2 &+& \\
\nonumber && \hskip -1.in + 4 ~ \Bigg{(} \frac{\delta
f_\mathrm{0}}{f_\mathrm{0}} \Bigg{)}^2 + 16 ~ \Bigg{(} \frac{\delta
d}{d} \Bigg{)}^2
\end{eqnarray}
Our goal is to achieve sub-percent
precision in amplitude calibration, therefore we need to keep the
relative uncertainties of every DFG parameter well below
$\simeq0.1\%$.

The currently accepted value of the gravitational constant, $G$, is
($(6.6742 \pm 0.0010) \times 10^{-11} \mathrm{m^{3}kg^{-1}s^{-2}}$).
This means, that there is a $\simeq0.015\%$ contribution to the
relative uncertainty on the induced displacement just by taking into
consideration the precision of previous $G$ measurements. $G$
contributes as the leading term in limiting the precision of
amplitude calibration if the uncertainties related to manufacturing
and/or measurement of the other parameters contributing to each of
the other four terms in equation~(\ref{errors}) is below 0.015\%.
Thus we require
\begin{eqnarray}\label{unc}
\frac{\delta m}{m} & \leq & 1.5 \times 10^{-4} \\ \nonumber
\frac{\delta r}{r} & \leq & 7.5 \times 10^{-5} \\ \nonumber
\frac{\delta f_\mathrm{0}}{f_\mathrm{0}} & \leq & 7.5 \times 10^{-5}
\\ \nonumber
\frac{\delta d}{d} & \leq & 3.75 \times 10^{-5} \\
\nonumber\end{eqnarray}
These levels of uncertainties adding up in
quadrature yield 0.035\% uncertainty in ($\delta x/x$), more than
adequate for a sub-percent amplitude calibration.

Considering the DFG described in sec.(\ref{idis}), with
$m_\mathrm{1}=m_\mathrm{2}=1.5\ \mathrm{kg}$,
$r_\mathrm{1}=r_\mathrm{2}=0.25\ \mathrm{m}$, rotation frequency of
$f_\mathrm{0}=51\ \mathrm{Hz}$ and distance from the TM of $d=2.5\
\mathrm{m}$, the required uncertainties in (\ref{unc}) translate to
\begin{eqnarray}\label{unc2}
\delta m = 2.25 \times 10^{-4} kg \\ \nonumber \delta r = 1.9\times
10^{-5} m \\ \nonumber \delta f_\mathrm{0} = 3.8 \times 10^{-3} Hz
\\ \nonumber \delta d = 9.4 \times 10^{-5} m
\\ \nonumber\end{eqnarray}
Most precision off-the-shelf balances can be used to measure the DFG
masses, while the ultimate precision of mass determination, $\delta
m$, is $\sim 50\ \mathrm{\mu g}$~\cite{balance2} with a state of the
art mass comparator. Uncertainties on $\delta r$ are determined by
machining precision and can be kept within $\sim 1\ \mathrm{\mu
m}$~\cite{cranfield}.

The uncertainty in the rotational frequency $f_\mathrm{0}$ can be
addressed by using a precision optical encoder to provide pulses
which can be used to phase lock the absolute angular position of the
DFG to an atomic clock or GPS. In this case, the uncertainty is
limited by the encoder itself or the servo system. For a rotation
period of $1/f_\mathrm{0}\ =\ 20\ \mathrm{ms}$ and an off-the-shelf
16-bit optical encoder providing a square pulse train at $3.2\
\mathrm{MHz}$ ($\simeq300\ \mathrm{ns/pulse}$), the relative
position of the square wave rising edge with respect to the atomic
clock signal can be determined for better than $\delta t \simeq 10\
\mathrm{ns}$. This allows for a high precision of $\delta
f_\mathrm{0}/f_\mathrm{0} \sim 10^{-6}$.

Distance $d$ could change somewhat over time when a DFG is used as a
calibration device. The thermal variations in the TM and DFG
housings are kept within fractions of a degree and should not play a
significant role. The tidal-compensation system, a servo-mechanism
acting on the position of the TM to compensate for earth-moon and
earth-sun tidal effects, displaces the TM locally with peak to peak
excursions of the order of $\sim 300\ \mathrm{\mu m}$ (see
\cite{tidalmodel} and \cite{tideLuca}). This kind of excursion can
be taken into account during the calibration.

Distance $d$ can be directly measured via laser based range finding
(i.e. Light Detection and Ranging, LIDAR) technologies, which can
provide better than $\delta d \simeq 1\ \mathrm{\mu m}$ uncertainty
in lab environments \cite{KaoruIEEE_2005}.

When direct distance measurement between the TM and the DFG is not
possible, an alternative method for finding $d$ can be adopted. The
$2 \omega_\mathrm{0}$ component can be measured as a function of $d$
by varying the DFG's position by a well known amount and using a
$\chi^2$ minimization procedure to estimate the effective distance
$d$. For simplicity, let the distance vary linearly
\begin{equation}
d(t) = d_\mathrm{0} + v ~ t
\end{equation}
where $v$ is the DFG's pivot velocity along the beam axis. Following
eq.(\ref{displ1}), the uncalibrated interferometer response
$R_{IFO}$ to the DFG's stimulus can be described as
\begin{equation}
R_{IFO} = \frac{K}{(d_\mathrm{0} + v ~ t )^4}
\end{equation}
where $K$ and $d_\mathrm{0}$ are free parameters. A linear sweep of the
pivot's position would provide an estimate of $d_\mathrm{0}$ while any
residual would provide information on any $d(t)$ component that
could potentially be significant. The uncertainty in $d_\mathrm{0}$ will be
statistical in nature and eventually will be limited to the
systematic uncertainty of the other parameters, such as the dipole
moment and the rotation frequency. In this case
\begin{equation}
\frac{\delta d}{d} \sim \frac{\delta r}{r} \sim 10^{-5}
\end{equation}

There are also other uncertainties that need to be addressed for
realistic measurements, most of them are second order in nature. For
example, stress under operation conditions results in the
deformation of the rotating DFG. The length change for a titanium 50
cm long 10 cm diameter rod holding two 1.5 kg masses at both ends is
estimated to be at the order of $10\ \mathrm{\mu m}$. For the
proposed DFG design (\ref{rotormodel}) this source of uncertainty
should be significantly less and can be carefully modeled, measured
and taken into account with a sub-$\mathrm{\mu m}$ accuracy.

An accurate alignment of the DFG is also necessary: the effective
arm length $\tilde{r}$ is altered if the plane of rotation of the
DFG is not aligned with the plane of the interferometer. Restricting
this change to $19\ \mathrm{\mu m}$ (same as the uncertainty
required for $r$) restrains the leveling of the DFG to
$0.7^\mathrm{\circ}$, which is achievable with commercial optical
positioning methods.

The absolute phase of the rotating DFG can be measured by phase
locking the DFG to an atomic clock or GPS. The phase uncertainty due
to $\delta t/t$ is therefore based on $\delta
f_\mathrm{0}/f_\mathrm{0} \sim 10^{-6}$, therefore the precision of
phase calibration for a perfectly oriented DFG can even be better
than $\simeq0.01\%$.

Placing the DFG out of line with the Fabry-Perot arm introduces
other second-order error sources. First, it creates a distance
$\tilde{d}$ which differs from $d$. Requiring their relative change
$(\tilde{d}-d)/d$ to be of the order of $10^{-5}$ sets an alignment
requirement to the cavity with an order of $1\ \mathrm{cm}$.
Additionally, a DFG not centered on the axis of the laser beam
introduces an error in phase determination. In order to achieve
0.01\% phase calibration this alignment requirement is constrained
to $250\ \mathrm{\mu m}$, which is still achievable with optical
positioning.

The quoted accuracy of calibration for the LIGO detector for recent
science runs~\cite{CalReview} is at the 6-10\% level and valid for a
broad range of frequencies and for the entire length of the science
run. The inherent accuracy of the calibration method itself is at
the order of 1-2\%~\cite{landry}. Using DFG as calibration tool this
can be pushed down to the subpercent level for amplitude and phase
calibration.

To take full advantage of this proposed calibration method for
interferometric GW detectors, we envision a DFG positioned at around
2-3 m from each end mirror of the two arms of the interferometer.
The rotation frequencies can be chosen such that subpercent level
calibration could be provided for the most sensitive region of the
detector response. The employment of two separate DFGs, rotating at
slightly different frequencies, would allow the calibration of the
two interferometer arms separately in a spectrally similar region.
Additionally, with longer integration times higher order harmonics
become detectable. Thus the device can be used for calibration of
interferometer response of frequency regions at points separated by
the DFG's rotation frequency. From signals at the higher harmonics,
information on the actual DFG parameters might also be deduced.

\section{Mitigation of Spurious Couplings from the DFG's Motor}
\label{couplings}

In interferometric GW detectors, using DFGs as a calibration tool
means that the new device will be put in close proximity (e.g. 2.5
meters) of the test mass for a prolonged period of time, while the
GW detector itself is in a continuous data taking mode. Thus, it is
necessary that spurious coupling of the DFG to the suspended mirror
be negligible, as detailed in this chapter. The only acceptable
effect on the GW data should be the fine and easily filterable lines
at the multiples of the rotational frequency of the DFG. Of most
concern is the electro-magnetic coupling via the motor driving the
system, the acoustic coupling via the local interferometer optical
sensors and the seismic vibrations induced by an unbalanced DFG.

\subsection{Electro-Magnetic Coupling}

There are two ways the motor's electro-magnetic field could couple
to the test mass. One coupling is the interaction of the motor's
electro-magnetic (EM) field with the interferometer electronics
residing next to the DFG. The other way is through the coupling of
the DFG's EM field with the coil-magnet system needed to drive the
TM in position.

With the proper Electromagnetic Interference (EMI) shielding in
place and using DC permanent magnet servo motors the parasitic
emission can be mitigated. The DFG could be equipped with a
non-integer gear ratio to completely separate the EM harmonics from
the Newtonian signal since the induced displacement appears at
harmonics of the rotation frequency of the DFG and not of the motor.

It is also possible to completely eliminate the mechanical coupling
via an Eddy Current Motor, which simplifies the DFG balancing and
bearing design. Alternatively one can use an air motor which also
eliminates the need for a gear-box mechanism.

\subsection{Acoustic coupling}

For the LIGO interferometers, acoustic signals near the detector
could potentially couple directly to the gravitational wave channel.
A possible coupling mechanism could consist of an acoustic stimulus
exciting the beam position on an optical sensor. If the sensor in
question is used to feedback on TM positions, the acoustic
excitation finds its way into the detector. This effect is mitigated
by installing the DFG in its own vacuum envelope.

\subsection{Seismic coupling}

One should also estimate the level of contamination into the GW
datastream, due to the coupling of seismic disturbances through the
ground, caused by the rotating device. This effect is the greatest
at the rotation frequency and should be considerably smaller at the
second and higher harmonics. For an ideally symmetric DFG, as
described in earlier sections, the dipole moment vanishes and so
does its contribution to the Newtonian field. Any asymmetry in the
system creates a non-null dipole moment at the rotation frequency,
introducing ground vibration. In this section we use a simple model
to estimate this cross-coupling for the initial LIGO case.

For an asymmetric DFG, the device's center of rotation will be
subjected to a sinusoidal force $F'$ at the rotation frequency
$\omega_\mathrm{0}$ whose RMS value along the beam axis can be
written as
\begin{equation}
F_\mathrm{rms}' = \frac{1}{\sqrt{2}} ~ \omega_\mathrm{0}^2 ~ \mathcal{M}_\mathrm{1}
\end{equation}
where $\mathcal{M}_\mathrm{1}$ is the dipole moment of the DFG. The
displacement $\delta x_\mathrm{react}$ of the reaction mass due to
the asymmetry, to first order approximation, is
\begin{equation}
\delta x_\mathrm{react} = \frac{m \, r}{\sqrt{2} \,
M_\mathrm{react}} ~ (\epsilon_r+\epsilon_m)
\end{equation}
where $\epsilon_r = \delta r / r$ and $\epsilon_m = \delta m / m$.
The TM displacement can then be expressed as
\begin{equation}
\delta x_\mathrm{rms} = \delta x_\mathrm{react} ~ \mathrm{R(f)}
\end{equation}
where $\mathrm{R(f)}$ is the attenuation factor provided by LIGO's
seismic isolation stage and suspension.

To estimate the motion of the cement slab beneath both the DFG and
the TM we select achievable uncertainty requirements of (\ref{unc}).
For a plausible reaction mass $M_\mathrm{react}$ of 100 tons
(assuming a concrete slab $10m \times 10m \times 0.5m$) its mass
displacement is
\begin{equation}
\delta x_\mathrm{react} = 6 \times 10^{-10} \mathrm{m}
\end{equation}
LIGO's stack\cite{stack} reduces this displacement down by a factor
of $\sim 10^6$ at $51\ \mathrm{Hz}$ while the suspension stage
\cite{TMsuspension} brings it down by an another factor of $\sim
(51\ \mathrm{Hz}/0.74\ \mathrm{Hz})^2 =4500$. This results in a TM
displacement of
\begin{equation}\label{xrmssc}
\delta x_\mathrm{rms} = 1.3 \times 10^{-19} m
\end{equation}
which is below the noise floor of LIGO and is only detectable with
$SNR = 3$ after ~half an hour integration time.

The above estimated effect of seismic coupling can be further
reduced by attaching the rotating DFG to a light slab with very
small reaction mass $M_\mathrm{react}$. The seismic signal of a
high-precision seismometer coupled to the slab, resulting from
system asymmetries, can be substantially reduced by iterative
adjustment of the balancing of the DFG. Attaching this balanced DFG
to a heavy slab with higher $M_\mathrm{react}$ will reduce $\delta
x_\mathrm{react}$ to be well below the ambient seismic field. The
reduction factor is given by the ratio of the reaction mass of the
light slab to the reaction mass of the heavy slab. This can lead to
a TM RMS displacement even orders of magnitude smaller than as given
in eq.(\ref{xrmssc}).

\section{Safety}

Significant kinetic energy (i.e. tens of kJs) is stored in the DFG
once it rotates and crucial safety considerations must be addressed.
There are two major points of failure management to be concerned
with. (a.) The vacuum chamber of the DFGs must be made strong enough
to withstand the damage of an accidentally disintegrating disk. This
is the standard solution for high speed gyroscopes. (b.) For added
security, the gap between the inner wall of the vacuum chamber and
the outer edge of the rotating disk must be kept relatively {\it
small}. In the event of an incident where the DFG's material starts
to yield or its angular acceleration is uncontrolled the disk will
expand radially touching the sidewall and slowly stop, preempting a
catastrophic failure. These conditions can be met using Finite
Element Analysis (FEA) aided design, in-house destructive testing of
sacrificial parts and relying only on X-ray rated base materials.

\section{Conclusion}

These initial feasibility studies of simple DFGs indicate that they
are capable of dynamically changing the local gravitational field by
an amount detectable by current interferometric gravitational wave
detectors.

The DFGs can be designed, manufactured, tuned and characterized to
be symmetric and safe enough to eliminate concerns about vibrations
and spurious couplings, once positioned in the proximity of one of
the suspended TMs.

The generated gravity gradient signal is proportional to the DFG's
quadrupole moment with its signature appearing at twice the rotation
frequency. At the present detector sensitivity level of LIGO,
systematic uncertainties due to the DFGs can be well below the
$0.1\%$ level in amplitude with insignificant timing uncertainties.
This apparatus provides a detector-independent calibration technique
that can significantly surpass the achievable precision of other
existing calibration methods.

The DFG also offers a unique and distinctive way to generate a
differential arm length displacement for gravitational wave
detectors. Apart from calibration objectives, it could also be used
to validate the expected noise generation and coupling mechanism of
Newtonian noise, possibly a limiting factor in advanced
gravitational wave detectors.

There are many details that need attention when designing and
manufacturing a practical device. Finite element analysis of the
DFGs and subsequent experimental studies are necessary to completely
understand the stresses the DFG is subjected to. The DFGs will be
enclosed in a separate vacuum chamber. A prototype design and test
will be necessary to balance the disk and test vibration control.
Other mostly practical problems, such as safety, can also be solved
as was shown in past applications/experiments that have used rapidly
rotating instruments.

\section{Acknowledgements}

The authors are grateful for the support of the United States
National Science Foundation under cooperative agreement PHY-04-57528
and Columbia University in the City of New York. We greatly
appreciate the support of LIGO collaboration. We are indebted to
many of our colleagues for frequent and fruitful discussion, and for
the LIGO Scientific Collaboration Review. In particular we'd like to
thank G.Giordano, R.Adhikari, V.Sandberg, M.Landry, P.Sutton,
P.Shawhan, D. Sigg, R.DeSalvo, H.Yamamoto, Y. Aso and C.Matone for
their valuable comments on the manuscript.

The authors gratefully acknowledge the support of the United States
National Science Foundation for the construction and operation of
the LIGO Laboratory and the Particle Physics and Astronomy Research
Council of the United Kingdom, the Max-Planck-Society and the State
of Niedersachsen/Germany for support of the construction and
operation of the GEO600 detector. The authors also gratefully
acknowledge the support of the research by these agencies and by the
Australian Research Council, the Natural Sciences and Engineering
Research Council of Canada, the Council of Scientific and Industrial
Research of India, the Department of Science and Technology of
India, the Spanish Ministerio de Educacion y Ciencia, The National
Aeronautics and Space Administration, the John Simon Guggenheim
Foundation, the Alexander von Humboldt Foundation, the Leverhulme
Trust, the David and Lucile Packard Foundation, the Research
Corporation, and the Alfred P. Sloan Foundation.

The LIGO Observatories were constructed by the California Institute
of Technology and Massachusetts Institute of Technology with funding
from the National Science Foundation under cooperative agreement
PHY-9210038. The LIGO Laboratory operates under cooperative
agreement PHY-0107417.

This paper has been assigned LIGO Document Number LIGO-P060056-00-Z.

\section{References}

\bibliographystyle{unsrt}
\bibliography{References_RotorPaper}

\end{document}